# Enhanced x-ray emission arising from laser-plasma confinement by a strong transverse magnetic field

E.D. Filippov[1,2], S.S. Makarov[2,3], K.F. Burdonov[1,4], W. Yao[4,5], G. Revet[1,4], J. Béard[6], S. Bolaños[4], S.N. Chen[7], A. Guediche[4], J. Hare[8], D. Romanovsky[1], I.Yu. Skobelev[2,9], M. Starodubtsev[1], A. Ciardi[5], S.A. Pikuz[2,9] and J. Fuchs[1,4]

[1] Institute of Applied Physics, RAS, 46 Ulyanov Street, 603950, Nizhny Novgorod, Russia

[2] Joint Institute for High Temperatures, RAS, 125412, Moscow, Russia.

[3] Department of Physics of Accelerators and Radiation Medicine Faculty of Physics, Lomonosov Moscow State University, Moscow, 119991, Russia

[4] LULI - CNRS, CEA, UPMC Univ Paris 06 : Sorbonne Université, Ecole Polytechnique, Institut Polytechnique de Paris - F-91128 Palaiseau cedex, France

[5] Sorbonne Université, Observatoire de Paris, PSL Research University, LERMA, CNRS UMR 8112, F-75005, Paris, France

[6] LNCMI, UPR 3228, CNRS-UGA-UPS-INSA, Toulouse 31400, France

[7] ELI-NP," Horia Hulubei" National Institute for Physics and Nuclear Engineering, 30 Reactorului Street, RO-077125, Bucharest-Magurele, Romania

[8] Imperial College, London, U.K.

[9] National Research Nuclear University "MEPhI", 115409, Moscow, Russia.

E-mail: edfilippov@ihed.ras.ru

**Abstract**

We analyze, using experiments and 3D MHD numerical simulations, the dynamics and radiative properties of a plasma ablated by a laser (1 ns, $10^{12}$-$10^{13}$ W/cm$^2$) from a solid target, as it expands into a homogeneous, strong magnetic field (up to 30 T) transverse to its main expansion axis. We find that as soon as 2 ns after the start of the expansion, the plasma becomes constrained by the magnetic field. As the magnetic field strength is increased, more plasma is confined close to the target and is heated by magnetic compression. We also observe a dense slab that rapidly expands into vacuum after ~ 8 ns; however, this slab contains only ~ 2 % of the total plasma. As a result of the higher density and increased heating of the confined plasma, there is a net enhancement of the total x-ray emissivity induced by the magnetization.

The investigation of the dynamics of strongly magnetized high-energy-density (HED) plasmas is a topic that has been recently the subject of significant effort by many groups. Permitted by the advent of new experimental facilities capable of developing strong magnetic fields [1–3], such investigations have led to major progress in diverse fields such as laboratory astrophysics [4–11] or inertial confinement fusion (ICF). In ICF, it has been realized that, by constraining the ion trajectories in the compressed core, and thus increasing their collision rate, magnetization increases the fuel ion temperature [12]. Magnetization has also been shown to localize heat transport [13], which improves laser beam propagation [14] and should reduce the growth of Laser-Plasma-Instabilities (LPI). Moreover, the presence of a strong magnetic field could substantially reduce the growth of hydrodynamic instabilities [15–17], and thus enhance fuel compression, as well as the heat conductivity within the fusion target. Overall, all of this would have the major benefit that, by easing the requirement for ignition, lower-cost laser drivers would be needed than in the case of traditional ICF schemes [18]. However, many questions pertaining to HED plasma dynamics in strong magnetic field still need to be addressed in order to properly design magnetized ICF experiments.

In this Letter, we investigate, using experiments and 3D MHD numerical simulations, the hydrodynamics and radiative properties of a plasma expanding from a solid plane, and the impact on it of applying a strong (up to 30 T), homogeneous and steady magnetic field. This is done in a configuration where, as shown in Fig.1 (a), the magnetic (B) field is oriented parallel to the surface of the target, which can represent a direct-drive ICF target outer surface or an indirect-drive ICF holhraum wall [19, 20]. This configuration has been experimentally investigated for many years [21, 22] including in more recent studies [23–25]. These focused mainly on the overall plasma dynamics over large distances, revealing significant instabilities which affect the long-range plasma propagation. Here we will focus on the dynamics of the plasma expansion near the wall itself. Consistently with previous studies, we observe that in its initial launching at the target surface, the plasma has a ratio $\beta = \frac{p_{plasma}}{p_{magnetic\ field}} > 1$ (where $p_{plasma}$ is overall pressure, i.e. the sum of the ram and thermal pressures), hence is able to expand rapidly. However, as the plasma expands and cools down, $\beta$ becomes $< 1$ at its leading edge, allowing the magnetic field to compress the plasma to form a cavity. This takes place already within 2 ns after the start of the expansion, i.e. over very short time scale compared to that typical of ICF compression experiment [18]. The compression is most efficient in the plane transverse to the field, and so the cavity emerges after $\sim 8$ ns as a thin slab which propagates away from the target. However, this slab represents only a small fraction (~2 %) of the total plasma. Beyond this global behavior, our detailed analysis reveals that the magnetic field induces a strong separation between the two flows (a slow, dense one and a fast, low-density one) that are observed to stem from the wall. Moreover, the plasma is heated as it is compressed and focused onto the axis in one dimension. Overall, as a combined result of higher density and increased heating, we find that the plasma emissivity in the x-ray region is enhanced by at least 50 % by the 30 T magnetization, compared to that of the unmagnetized plasma.

The experiment, which is shown in Fig.1 (a), was conducted on the ELFIE (LULI, Ecole Polytechnique, France) and TITAN (LLNL, USA) laser facilities. A high-power laser (see caption of Fig.1 for details) irradiates a solid $CF_2$ planar target, from which a plasma is ablated and expands into vacuum (along the axis X) as well as the large-scale magnetic field, when it is applied. The magnetic field, which is directed along the axis Z, is generated by a modified Helmholtz coil [2]. It can be adjusted between 0-30 T on different shots, and can be considered, with respect to the scales of the expanding plasma, steady and homogeneous [26]. The plasma is diagnosed in several ways. First, the emission of the plasma in the X-ray domain is collected and analyzed using two time integrated instruments (see also the Methods section of the Supplementary Information): a focusing spectrometer with spatial resolution (FSSR) [27] and a variable line spaced grating spectrometer (VSG) [28]. The two are complementary: the VSG has a lower spectral resolution, but a better sensitivity, than the FSSR. An image obtained by the FSSR of the plasma in the X-Y plane is shown in Fig.1 (b); the complementary image obtained in the X-Z plane is shown in Supplementary Fig.1. Second, an auxiliary optical laser probe is directed along the axis Z. Using an interferometry setup (we use the same setup as detailed in Ref. [6]), we measure with temporal resolution the low-density part of the plasma, as shown in Fig.1 (c-e).

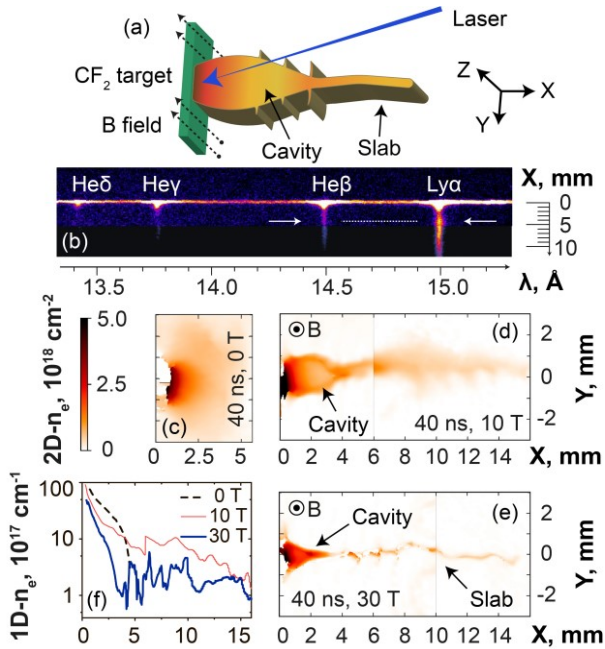

*Figure 1: (a) Setup of the experiment. A solid (Teflon, $CF_2$) target, mimicking an ICF hohlraum wall, is immersed in a large-scale, steady, axial and strong (30 T) magnetic field (sketched by the dashed lines) that is aligned with the target surface. A long (0.6 ns FWHM) high-power (40 J, $4 \cdot 10^{13}$ W/cm$^2$) laser pulse $\lambda = 1$ μm irradiates the wall, inducing plasma heating and expansion. The laser was focused through a 2.2 m focal length lens (f/21) and a random phase plate [29]. (b) image of the X-ray emission analyzed by the FSSR spectrometer of the laser-induced plasma expansion inside the transverse magnetic field of 30 T strength. What is seen is the spectrally resolved image of the plasma in the XY plane. Arrows demonstrate the increase in emissivity in spatial region ∼ 4 mm. (c-e) 2D maps of the plasma electron density 40 ns after the laser irradiation on target (located on the left edge of the images), for various strengths of the applied external magnetic field, as indicated. The color scale shown in (c) applies for all images. The images in panels (d) and (e) are, as panel (a) reconstructed from two different shots. (f) Corresponding radially integrated electron density as a function of the distance from the target, i.e. the 1D densities are obtained from 2D maps as shown in (c-e) and integrated over the axis Y. The lines correspond, respectively, to the case of a free expansion (black, dashed), 10 (thin red) and 30 T (thick blue).*

The impact of applying a transverse external magnetic field of increasing strength on the plasma propagation is shown in Fig.1 (c-f), as retrieved by optical probing of the plasma, and Fig.2, as retrieved from collecting the x-ray emission of the plasma. We briefly recall here the features of the unmagnetized plasma, i.e. a largely divergent expansion along the X-axis (see Fig.1 (c)) in which the plasma also cools down rapidly (see Fig.2 (d)). These features are further detailed in Ref. [26] (see in particular Figs.4 and 10 of that paper). In stark contrast, when applying the external magnetic field, we observe that the plasma is focused on axis by the magnetic tension into a conical shape, from which emerges a slab [25] which is observed by the optical (Fig.1 (d-e)) and x-ray diagnostics (see Fig.1 (b) and Supplementary Fig.1). From early snapshots of the interferometry diagnostics (not shown), we observe that the cavity already starts to form at ∼ 2 ns, and that the slab emerges from the cavity at ∼ 8 ns. The slab is formed as the plasma is able to expand along the magnetic field lines (in the XZ plane), but is highly compressed and modulated [25] in the YX plane (see Fig.1 (d-e)). This global morphology is consistent with previous observations [23–25]. Extending these previous studies, we can observe in Fig.1 (d-e) that the plasma compression in the YX plane increases with the magnetic field strength. At the same time, there is a noticeable decrease of the overall plasma mass flow in the region $x > 2$ mm that can be observed in Fig.1 (f). Note that the optical probe is limited to measuring the plasma in the region $x > 1$ mm since beyond this, the optical probe is refracted.

As the plasma flow is quenched by the magnetic field far away from the target, we would expect that this is linked to plasma being retained against the target. This is actually what is observed by the X-ray diagnostics, which ideally complements the optical probing since they can resolve the plasma emission down to the target surface. Moreover, since the x-ray ion emission spectra are time-integrated, they offer the advantage of informing us on the global modification of the plasma flow under the influence of the external magnetic field, whereas the optical probing is limited to snapshots of the plasma evolution. Fig.2 (a) compares the relative intensity of the plasma emission recorded in a broad x-ray energy range and for different strengths of the applied magnetic field, as recorded by the VSG spectrometer. We readily observe a much more constrained plasma emission profiles when increasing the magnetic field from 20 (red) to 30 T (blue), respectively. This can be observed consistently as well in the emission recorded by the FSSR (see Fig.2 (b)), which, due to its higher resolution allows us to further analyze separate plasma emission spectral lines (see Fig.1 (b)). In all spectral lines (as illustrated in Fig.2 (b) by the case of the $Ly_\alpha$ line), we clearly see that the relative intensity of the emission drops more steeply as the magnetic field strength is increased. We also observe that the x-ray spectral intensity increases again in the region 2–5 mm from the target surface, and even exceeds the intensity of the freely propagating plasma farther away. As shown in Fig.1(e), integrating in space and time the overall x-ray emission as recorded by the broadband VSG measurements show that there is at least a 50 % of the emission when the plasma is magnetized at 30 T compared to the unmagnetized plasma.

*Figure 2: (a) Relative intensity profile measured by the VSG spectrometer in a wide spectral (integrated over 0.65-1 keV) and spatial range. (b) Relative intensity of the spectral line Lyα measured by the FSSR spectrometer. (c) Profile of the volumetric electron density inferred in the plasma, and along the axis X, from the FSSR data (see text). (d) Same as (c) for the plasma electron temperature. For all panels, three cases, corresponding to different strengths of the applied magnetic field, are shown: 0 (black, dashed), 20 (red, thin) and 30 (blue) T. Note that the values indicated in panels (c) and (d) are time-averaged due to time-integrated measurement performed by the FSSR spectrometer. (e) Evolution of the integrated plasma emission intensity (normalized to its value at B=30 T), integrated in space and time, and deduced from the intensity profiles shown in (a), as a function of the B-field strength.*

We will now discuss the plasma dynamics, as retrieved from the high-resolution x-ray data measured by the FSSR. Two techniques have been used. The first (detailed in the Methods section of the Supplementary Information) is to fit the relative intensity of the lines present in the spectra. This was done for multiple line ratios [30], and the inferred density was confirmed by the measurements of the interferometer. This yields the *time-averaged*, and *local (volumetric)* densities and temperatures in the plasmas. The results of this analysis, performed for various cases (unmagnetized and with 20 or 30 T applied), are shown in Fig.2 (c-d).

Obviously, very near the target surface, where the plasma ram pressure dominates the magnetic pressure, the plasma dynamics should be independent of the applied magnetic field. This is consistent with what we find, with the electron temperature being about 300 eV on the target surface in all cases (see Fig.2 (d)). Farther from the target, i.e. for $x > 3 - 4\ mm$, we observe that the plasma temperature and *volumetric* density both increase when the transverse magnetic field is applied. Note that the density increase is consistent with what we observe in Fig.1 (d-e) with the optical probing, i.e. it corresponds to the tip of the cavity, where the plasma becomes strongly compressed by the magnetic field and forms the slab: as a result of the compression, the *local* density in the slab indeed increases. However, as highlighted by Fig.1 (f), the global mass flow decreases with the increasing magnetic field. We also stress that the amount of plasma in the slab represents only ~ 2 % (as deduced from the ratio of the density close to the target to that in the slab, see Fig.1(f)) of all the volume of plasma.

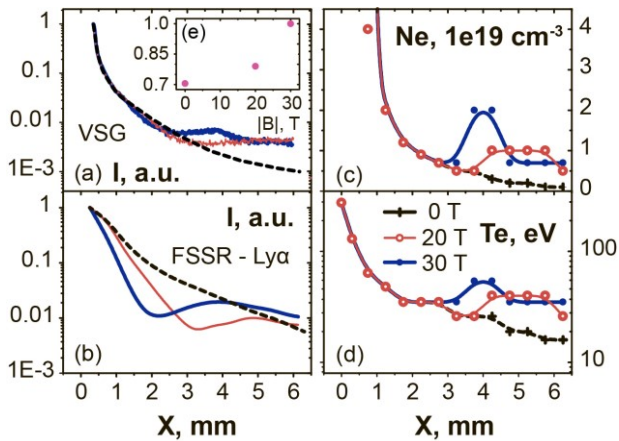

To go beyond this first time-integrated analysis of the X-ray spectra, and obtain a refined picture of the plasma evolution close to the target surface, we simulate the spatial profile of spectral lines by solving the system of kinetic equations (detailed in the Supplementary Information) governing the emission, and taking into account the processes of recombination and excitation in ions. This procedure was applied for two spectral lines of multicharged fluoride ions – $Ly_\alpha$ and $He_\beta$ – both for the unmagnetized and 30 T magnetized cases. The results are shown in Fig.3 for the $Ly_\alpha$ line, but similar results are obtained from the $He_\beta$ line (as shown in the Supplementary Information), showing the robustness of the analysis. To be able to correctly fit the spatial profile of the line along X, we find that we have to model the plasma as composed of two plasma fractions, each having different velocities. Of course, the ionic distribution of the plasma in terms of velocity is actually a continuous function. However, as shown below in Fig.3, we find that approximating such a continuous distribution function by a two-step, piece-wise function and therefore a two components plasma, in both in the unmagnetized and magnetized cases, leads to good fitting of the x-ray line spatial profile.

As shown in Fig.3, what we find is that the first plasma component, that extends from the target surface to several millimeters, is rather slow, but contains most of the plasma. It is characterized by velocities up to $4 \cdot 10^7$ cm/s. The second one that extends beyond 4 mm from the target surface is fast, with velocities $\sim 10^8$ cm/s, but concerns only a low-density part of the plasma, i.e. it contains only 5-10 % of the total number of particles. Note that these velocities are not thermal ones, but correspond to a ram motion of the plasma. They are consistent with those that can be derived from observing the progression of the plasma tip with the optical interferometer. Note also that when simulating independently the $Ly_\alpha$ and $He_\beta$ spectral lines, we obtained slightly different velocities for the fast and slow components (see Fig.3 of the Supplementary Information), which can be explained by the inhomogeneity of charge distribution in the plasma, i.e. the hotter plasma having a higher percentage of bare nuclei.

We observe in Fig.3 that the 30 T external magnetic field, when applied, has the notable effect of further separating the two inferred components. Compared to the unmagnetized case, most of the plasma finds itself confined by the magnetic field against the target, which translates into the averaged plasma velocity near the target becoming lower ($4 \times 10^7$ cm/s in the magnetized case vs $2.5 \times 10^7$ cm/s in the 30 T case). The fact that the slow, high-density and the fast, low-density components of the plasma are further separated in the presence of the external magnetic field can be easily understood. Indeed, the slow plasma component has an averaged velocity for ions with energies corresponding to the Larmor radius (about 0.3 mm) which does not exceed the transverse plasma size. This explains the increased plasma confinement by the magnetic field: the forward motion of these ions is slowed down by the magnetic field, which leads to an enhanced plasma emission near the target surface ($Z < 4$ mm). This is in contrast with the second plasma component for which the Larmor radius (~ 2 mm) exceeds the transverse plasma size. These ions are associated to the plasma emission which is recorded far from the target surface ($Z \gg 4$ mm).

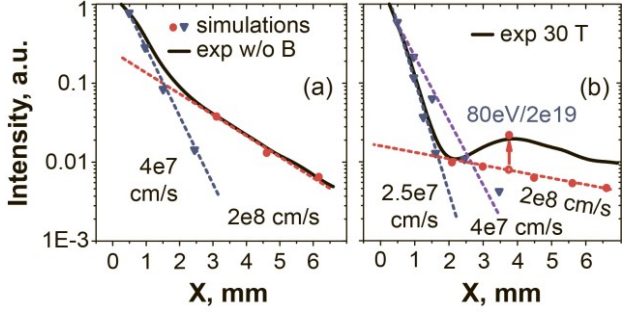

*Figure 3: (a) Experimental (full) and simulated (dashed) spatial profiles of the x-ray resonance line Lyα in the free expansion case. The theoretical intensities were normalized by the corresponding experimental points. (b) Same in the case where a transverse magnetic field $B = 30\,T$ is applied. In the simulations, the excitation processes, as well as the recombination ones, were taken into account, as detailed in the Supplementary Information. Note that the modelling, for the same velocity of the plasma component, but different densities can result in different slopes for the spectral intensity.*

The dynamics of the plasma confinement, retrieved from the experimental results detailed above, is further supported by 3D MHD simulations of the process performed using the GORGON code (see Supplementary Information for details about the simulations). The results of these simulations, run in the case without magnetic field and with a 30 T magnetic field, are shown in Fig.4. We can observe in Fig.4 (a-b) that the compression of a thin slab also occurs in the simulation, and that this compression leads to increased density (see also Fig.4 (d)) and temperature along the axis. This, in turns, results in higher emissivity of the plasma (integrated from 150 eV to 1100 eV) in the magnetized case (see also Fig.4 (c)). Overall, integrating the radiation in space, time and over the 150-1100 eV spectral range, we find that the confinement induced by the B-field drives the emission of the plasma as the B-field increases (by a factor ∼ 6, passing from 0 to 30 T), as can be seen in Fig.4 (e). The larger increase recorded in the simulation compared to that measured by the VSG and shown in Fig.1 (e) could be due to the limits of the modelling of the x-ray emission.

In summary, our experiment demonstrates the ability to strongly impact the dynamics of a laser-driven plasma flow from a solid wall and in vacuum by applying an external transverse magnetic field. In particular, we have shown that the magnetic field leads to a reduction of the overall plasma flow, with a more precise analysis revealing the increased separation of the plasma into two components: a dense, slow one, increasingly confined against the target as the magnetic field increases, and a low-density, fast one, which is redirected on axis by the magnetic field. The compression applied onto the plasma by the magnetic field in the plane transverse to the latter induces an increase of the local plasma density and temperature, both of which drive an increase in the plasma x-ray emission. Both of these factors, an increased emission, and an increased confinement of the dense plasma could constitute, on top of enhanced heating in the fuel, an additional interest for magnetizing ICF targets. However, the fact that the hotter fast plasma component, redirected by the magnetic field, could actually reach the ignition capsule earlier than in the unmagnetized case needs to be considered in the overall design of indirect-drive magnetized ICF.

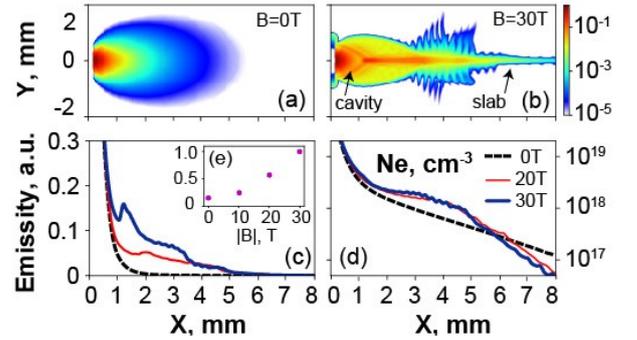

*Figure 4: 3D MHD simulation results. (a)-(b) Pseudocolor maps of the relative plasma emission intensity (from 150 eV to 1100 eV) in the XY plane, in logarithm, and for (as indicated) $B = 0\,T$ and $B = 30\,T$, respectively. (c) Profiles of relative emission intensity, integrated over time (up to 24 ns, which is the maximum time in the simulation) and over the Y and Z-directions, for $B = 0$ (black, dashed), 20 (red, thin) and 30 T (blue) (similarly as in Fig.2 (a)). (d) Profiles of*

*electron density in logarithm scale, integrated over time and over the Y and Z-directions, for $B = 0$ (black, dashed), 20 (red, thin) and 30 T (blue). (e) Evolution of the integrated plasma emission intensity (normalized to its value at B=30 T), integrated in space and time, as a function of the B-field strength.*

The platform used for the present study could also be relevant to astrophysical investigations. Plasma interaction with crossed magnetic field is indeed a phenomenon taking place in a wide variety of astrophysical environments, like at the edge of disks surrounding forming young stars, or when coronal mass ejections (CMEs) [31] propagate away from a star [32]. Notably, the ability observed here for plasma to propagate against a transverse magnetic field could be of interest to disk edge physics: by allowing removal of matter from the disk and across the gap separating the disk from the star, it could participate to the postulated mechanism by Shakura and Sunyaev [33] of disk slowing down.

### Acknowledgments

The reported study was funded by RFBR, project number 19-32-60008. This work was supported by funding from the European Research Council (ERC) under the European Unions Horizon 2020 research and innovation program (Grant Agreement No. 787539). Part of the experimental system is covered by a patent (n 1000183285, 2013, INPI-France). The research leading to these results is supported by Extreme Light Infrastructure Nuclear Physics (ELI-NP) Phase II, a project co-financed by the Romanian Government and European Union through the European Regional Development Fund.

## Supplementary Information

1. Methods

**FSSR X-ray spectrometer:** The FSSR is a x-ray Focusing Spectrometer with Spatial Resolution which uses a spherically bent mica crystal with 2d = 19.9149 Å and curvature radius of R=150 mm for the mentioned experiment. This spectrometer allowed to observe plasma ion emission in the spectral range 13-16 Å with a spatial resolution about 0.1 mm along the X-axis and a spectral resolution better than $\lambda/d\lambda = 1000$. Multicharged fluorine ions were analyzed through their hydrogen-like emission (transition 2p–1s) with dielectronic satellites and through the resonance series of helium-like ions (transitions 3p–1s, 4p–1s, 5p–1s etc.). Due to the presence of the magnetic field generating coil, the crystal was placed in a 2D scheme at a distance of 480 mm from the plasma source and the image plane was on the Rowland circle. Spectra were recorded using Fujifilm Image Plates of type TR, which were placed in a cassette holder protected from the optical radiation. The main spatial resolution of the spectrometer is along the X-axis. It is however also able to spatially resolve the plasma along the other axis (Y or Z in the experiment). Thus, the FSSR was able to observe the dynamics of the plasma expansion (see Supplementary Fig.1 and Fig.1 (b) of the main paper) which fully consistent with what is observed by the optical interferometry probing (see Fig.1 (c-f) and [25]). The image in Fig.1 (b) of the main paper was constructed from two different laser shots, the first being with the target located at the edge of the coil, in the field of view of the spectrometers, the second being with the target shifted within the coil, so that the plume farther from the target could be analyzed. This is possible since, due to the high reproducibility of the magnetic field generation, the expanding plasma dynamics is extremely reproducible [26].

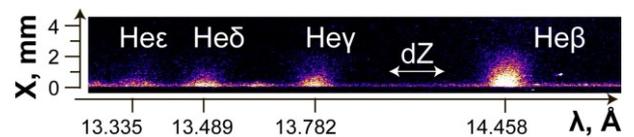

*Supplementary Figure 1: FSSR x-ray raw image showing the same spectral lines as in Fig.1 (b) but*

*when probing the plasma along a different line of sight, here along the Y-axis of plasma expansion (see Fig.1 (a)). Hence w*hat is seen is the spectrally resolved image of the plasma in the XZ plane. *The absence of the Ly$_\alpha$ line here is due to the different target-to-crystal distance that was changed to 214 mm in order to better observe the 2D dynamics (with higher spatial resolution along Z axis and magnification $m = 1$ in meridional plane) of the plasma expansion.*

To retrieve the plasma parameters from the spectra recorded by the FSSR, the following procedure is applied. Near the target surface, i.e. for x < 0.5 mm, the relative intensities of the Ly$_\alpha$ line and of its dielectronic satellites were simulated using the radiative-collisional atomic code PrismSpect [34] with a steady-state kinetic model. Farther from the target, the plasma is in a recombining mode [35], so the quasi-stationary kinetic approach [30] was rather used to retrieve the electron temperature and density profiles from the spectra. For this purpose, the relative intensities of He-like lines of fluorine (transitions 1s3p-1s$^2$, 1s4p-1s$^2$ etc.) were used [36].

**VSG X-ray spectrometer:** The variable line spaced grating spectrometer (VSG) is an x-ray spectrometer that allows to investigate a wider spectral range – 200–2000 eV than the FSSR, but with a lesser spectral resolution [37]. It consists of a diffraction grating (1200 lpi in average), a spectrometer housing, and a removable nose cone that can house slits for spatially resolved measurements. In this experiment, the front of the housing was equipped with a vertical slit (i.e. aligned with the Y-axis), allowing to spatially resolve the plasma emission along the horizontal X-axis, which was the one of the plasma expansion. As for the FSSR, the image of the plasma was also recorded on a TR image plate, hence the measurement is time-integrated. In front of the image plate was a sheet of aluminized plastic to serve as a light-tight filter. For the deconvolution of the VSG data, the dependencies of the image plate sensitivity [38] and of the reflectivity of the grating [39] on the energy of photons were both taken into an account.

**Simulation of the experiment using the GORGON code:**

GORGON is a 3D resistive, single-fluid, bi-temperature and highly parallel magneto-hydrodynamic (MHD) code. It was originally developed to model Z-pinches and has been widely used in high-energy-density laboratory astrophysics researches [40, 41]. In the code, the plasma is assumed to be optically thin, the ionization process is obtained from an analytical approximation to an LTE Thomas-Fermi model, and the radiation emission is computed assuming recombination (free-bound) and Bremsstrahlung radiation [42] with an upper limiter given by the black body radiation rate (Stefan-Boltzmann's law).

The simulation box is defined by a uniform Cartesian grid of dimension 8mm x 8mm x 12mm and a number of cells equals to 400 x 400 x 600. The spatial resolution is homogeneous and its value is $dx = dy = dz = 20$ μm. The initial laser interaction with the solid target is performed using the DUED code [43], which solves the single-fluid, 3-Temperature equations in two-dimensional axisymmetric, cylindrical geometry in Lagrangian form. The code uses the material properties of a two-temperatures equation of state model (EOS) including solid state effects, and a multi-group flux-limited radiation transport module with tabulated opacities. The laser-plasma interaction is performed in the geometrical optics approximation including inverse-Bremsstrahlung absorption.

The laser and target parameters are taken to be similar to the experimental ones. At the end of the laser pulse (about 1 ns), the plasma profiles of density, momentum and temperature from the DUED simulations are remapped onto the 3D Cartesian grid of GORGON with a superimposed magnetic field and used as initial conditions. The purpose of this hand-off is to take advantage of the capability of the Lagrangian code to achieve very high resolution in modeling the laser-target interaction. The uniform magnetic field is aligned with the target surface (in the x-direction) and has magnitudes ranging from 10 T to 30 T. We consider "outflow" boundary conditions, where plasmas and electromagnetic fields will simply exit the simulation box. The computational vacuum cut-off density is set to $10^{-5}$ kg/m$^3$, under which the momentum, mass, energy and current densities are zero. To remove the initially imposed symmetry and to account for the effect of inhomogeneities in the laser intensity over the focal spot, we introduce uniformly distributed random perturbation on the plasma velocity components, with a maximum amplitude of ±5% the initial value. We note that this simulation method has been benchmarked in a variety of similar configurations [8, 26].

## 2. Simulation of the relative intensities of the x-ray spectral lines registered by the FSSR spectrometer.

Our second method to analyze the spatial profiles of the spectral lines recorded by the FSSR is detailed below. This approach is not only intrinsically refined compared to the first, time-averaged approach detailed above, but it allows also to better fit the spectral lines. Indeed, we can observe in Figure 2 (c-d) that the plasma density and temperature retrieved by the time-averaged approach varies monotonically close to the target surface (x < 3 mm) while obviously the intensity of the spectral lines (see Figure 2b, and the arrows in Fig.1.b indicating the increase in intensity of the resonance lines) present strong variation in the same region, especially at 30 T. The time-dependent approach developed below was found to solve these issues and thus retrieve refined plasma parameters from the measurements. In general, the absolute intensity of the spectral line caused by the transition from the excited level *n* to the level *k* is known to be as follows:

$$I_{nk} = E_{nk} \cdot A_{nk} \cdot N_n^z \qquad (1)$$

where $E_{nk}$ is the photon energy, $A_{nk}$ is a radiation probability, and $N_n^Z$ is the population of corresponding ions with charge $Z$ excited to the level n. The population of the excited states can be determined in a quasi-stationary approach by solving the system of kinetic equations and expressed through the population of ground states taking into account processes of recombination in ions with a charge Z+1 (see Supplementary Fig.2) and excitation in ions with charge Z [30]:

$$N_n^Z = \beta_n \cdot N_e \cdot N_1^{Z+1} + S_n \cdot N_1^Z \qquad (2)$$

where subscript 1 marks the ground state, $N_e$ is the electron density, $\beta_n$ and $S_n$ are population coefficients by the recombination and excitation, respectively. The population of the ground states is given by the following differential equations for the approximation of one-electron transitions:

$$\frac{dN_1^Z}{dt} = \beta^Z \cdot N_e^2 \cdot N_1^{Z+1} - (S^Z \cdot N_e + \beta^{Z-1} \cdot N_e^2) \cdot N_1^Z + S^{Z-1} \cdot N_e \cdot N_1^{Z-1} \qquad (3)$$

Here $\beta^Z$ and $S^Z$ are recombination and ionization rates for the processes $A_{Z+1}$ + 2e → $A_Z$ +e, and $A_Z$ +e → $A_{Z+1}$ +2e, respectively.

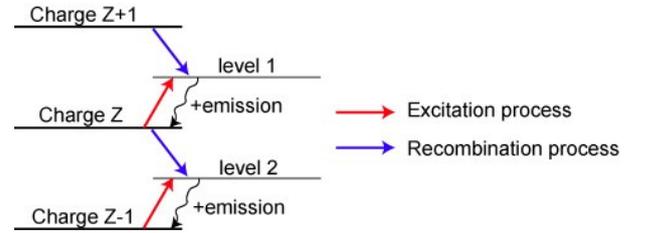

*Supplementary Figure 2: Scheme of the mechanisms producing the populations implied in the generation of the spectral resonance lines of multicharged ions. Here are depicted the levels with charges Z-1, Z, Z+1, each being in its ground states; levels 1 and 2 are excited levels. The red and blue lines correspond to the excitation or recombination population mechanism into the excited level. The further transition of an electron from each excited state (black arrows) leads to the ion emission of the corresponding spectral line.*

For the resonance line Ly$_\alpha$ (transition 2p-1s, Z corresponds here to H-like state) in the recombining plasma (i.e. the contribution of ionization is negligible), Equation (3) can be easily modified to the following:

$$I_{21}(t) = E_{21} \cdot A_{21} \cdot \beta_2(N_e, T_e) \cdot N_e \cdot N_1^{z+1}(0) \cdot exp\left(-\int_0^t \beta^z(N_e, T_e) N_e^2 dt\right) \quad (4)$$

where $N_1^{z+1}(0)$ is the initial value of a bare nuclei concentration at time $t = 0$. Here we consider the plasma to expand with a fixed velocity $v = x/t$, where $x$ is the distance from the target surface. It gives the opportunity to modify Eq. (4) with functions depending only on the distance, and to integrate them using the known dependences $N_e(x)$, $T_e(x)$, which are shown in Fig.2 (c, d) of the main paper. Since the plasma parameters measured in the experiment can be considered as piecewise functions which are constant at certain spatial interval, then the spectral line intensity profile can be expressed for each interval $m$ as an exponential function that has a dependence on the distance $x$ from the target:

$$I_{21}(x) = E_{21} \cdot A_{21} \cdot \beta_2(m) \cdot N_e(m) \cdot N_1^{z+1}(m-1) \cdot exp\left(\beta^z(m) \cdot N_e^2(m) \cdot (x_0(m) - x)/v\right) \quad (5)$$

where $x_0(m)$ is the initial coordinate for each spatial interval. Hence, by now varying the velocity of the hydrodynamic flow, we can simulate the relative intensity of the resonance line Ly$_\alpha$ and thus retrieve the parameters of the plasma expanding freely in a vacuum as well as across the magnetic lines. The result is shown in Fig.3 of the main paper.

Similarly, the intensity of spectral lines other than Ly$_\alpha$ can be derived as well starting from Equation (3). In order to verify the obtained data, the same technique was applied for the second intense spectral line in the investigated range – He$_\beta$, which measured spatial profiles of relative intensity, for the magnetized and unmagnetized cases, are shown in Supplementary Fig.3 (a). As a result of this procedure, the simulated relative intensity profiles, from which the plasma flow velocities of propagation are retrieved, are shown in Supplementary Fig.3 (b, c) for, respectively, the unmagnetized and magnetized cases. We note that the plasma dynamics inferred from simulating the He$_\beta$ spectral line is similar to that obtained by the Ly$_\alpha$ line (see Fig.3 of the main paper), which validates the robustness of the applied method. The slight differences in the retrieved plasma parameters, as inferred from the Ly$_\alpha$ and He$_\beta$ spectral lines, for the fast and slow components can be explained by the inhomogeneity of the charge distribution in the plasma, i.e. the hotter plasma having a higher percentage of bare nuclei.

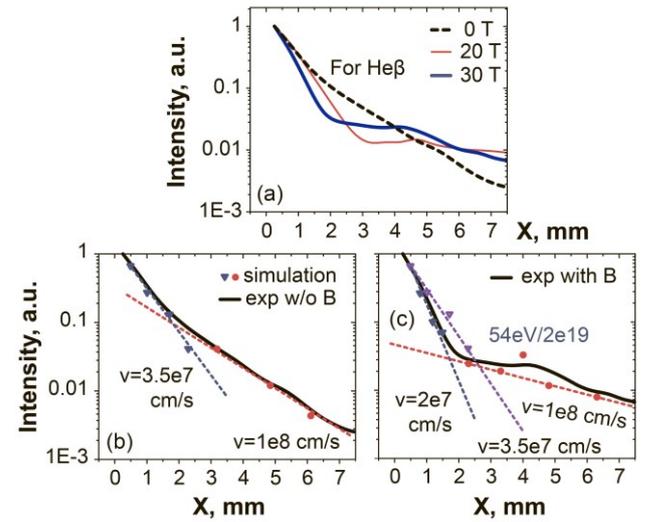

*Supplementary Figure 3: (a) Relative intensity of the spectral line He$_\beta$ measured by the FSSR spectrometer in the experiment both for a free expansion (black) and for the propagation in the transverse magnetic field with a strength of 20 (red) and 30 T (blue). (b) Experimental and simulated intensity profiles for the resonance line He$_\beta$ in the free expansion case. Theoretical intensities were normalized by the corresponding experimental point. The inferred plasma parameters are indicated. (c) Same as (b) but for the case when the transverse magnetic field B=30 T is applied.*

We should note that, for the Ly$_\alpha$ line and in the spatial region close to ~ 4 mm, i.e. in the region where the plasma is being refocused and shocked [25] by the magnetic field (see Fig.2 (c, d) of the main paper), it is not possible to fully simulate the spatial variation in the line intensity (see the red arrow in Fig.3 (b) of the main paper pointing to the discrepancy in amplitude between the simulated line intensity and the measured one). However, the simulated $He_\beta$ intensity profile fits the experimental line profile much more closely. It can be explained by the fact that Equation (4) implies the plasma to be purely recombining. However, an additional plasma heating in the region x ~ 4 mm (which is valid for a higher ionic state), as due to the local strong focusing of the plasma induced by the magnetic field (see Fig.4 (b)), can lead to the situation when the excitation of the ground state of the H-like ion by electron collisions brings in additional contribution to the emissivity of the $Ly_\alpha$ line. In order to take into account the contribution of such excitation processes into the ion population, we used the fact that the intensity of $He_\beta$ line due to the recombination (transitions 1s–1s3p) and the intensity of $Ly_\alpha$ line due to the excitation (transitions 1s–2p) are based on the same concentration of H-like fluoride ions (see Supplementary Fig.2 with $Z = 8$, level 1 which corresponds to level 2p, level 2 – 1s3p). Then, considering Equation (2), the intensity ratio of these lines can be evaluated as:

$$I(Ly_\alpha)/I(He_\beta) = E(Ly_\alpha)/E(He_\beta) \cdot A(Ly_\alpha)/A(He_\beta) \cdot S_2(N_e,T_e)/(\beta_2(N_e,T_e) \cdot N_e) \quad (6)$$

As a result, it gives the alternative measurement of the electron temperature about 80 eV at that location (close to ~ 4 mm) which corresponds to the measured intensity of the resonance line (shown in fig.3 (b)). Note that the real spectral ratio $I(Ly_\alpha)/I(He_\beta)$ for the emission attributed to the shocked ionizing plasma at a particular time could be several times higher than the observed time-integrated value, and so the value of $T_e \sim 80 \, eV$ should be considered as a lower estimation.

### 3. X-ray data for lower laser intensity (about $10^{12}$ W/cm$^2$).

To complement the results shown in the main paper (4·10$^{13}$ W/cm$^2$), we here report on complementary measurements conducted at lower laser intensity, and which concur with the results obtained at higher intensity. To do this, we increased the size of the focal spot to reduce the intensity to about 2x10$^{12}$ W/cm$^2$ (20 times less than the nominal intensity). The comparison between the cases with "high" and "low" laser intensities is presented in Supplementary Fig.4 for a plasma immersed into the magnetic field with 0-30 T strength and as measured by the VSG spectrometer after taking into account all instrumental functions. Obviously, the total emissivity of the plasma is decreased near the target in the low intensity case, however, the spatial profile of the emissivity (see Supplementary Fig.5 (a)) is similar to that recorded in the high intensity case (see Fig.2 (a)). Notably, as shown in Supplementary Fig.5 (c), we also observe in this reduced laser intensity case a net increase (by ~ 50 % as well) of the overall x-ray emission in the 30 T magnetized case compared to that of the unmagnetized case. The FSSR data is also demonstrating the same dynamics of the plasma expansion in the transverse magnetic field as in the higher intensity case (compare Supplementary Fig.5 (b) to Fig.2 (b)), although the intensity of the spectral lines is here quite low far from the target, which makes it difficult to measure the plasma parameters there.

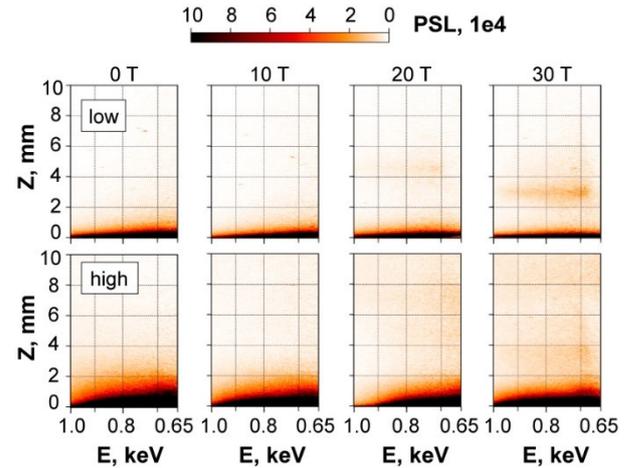

*Supplementary Figure 4: (top) VSG x-ray images (in PSL) for a laser-induced plasma immersed into a magnetic field of different strengths 0-30 T and when the laser intensity is about 2x10$^{12}$ W/cm$^2$ on*

*the target. (bottom) The same but for the laser intensity of 4x10$^{13}$ W/cm$^2$.*

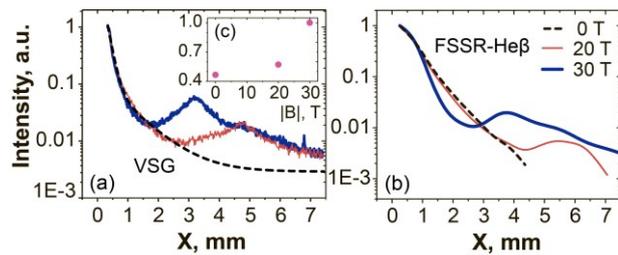

*Supplementary Figure 5: (a) Spatial profile of the emitted x-rays as recorded by the VSG and integrated over the 0.65-1 keV spectral range, for a 2x10$^{12}$ W/cm$^2$ laser intensity on target. The plasma was immersed into the magnetic field with strength of 30 T (blue), 20 T (red, thin) or expanded freely in vacuum (black dashed). (b) Same but for the He$_β$ spectral line measured by the FSSR spectrometer. (c) Evolution of the integrated plasma emission intensity (normalized to its value at B=30 T), integrated in space and time, and deduced from the intensity profiles shown in (a), as a function of the B-field strength.*